\documentclass[12pt,a4paper]{article}
\pdfoutput=1
\usepackage{jheppub}
\usepackage{amssymb}
\usepackage{mathrsfs}
\usepackage{slashed}
\usepackage{graphicx,color}
\usepackage{amsmath}
\usepackage[greek,english]{babel}
\usepackage{subcaption}
\usepackage[colorlinks=true,linkcolor=blue,citecolor=blue]{hyperref}%

\begin{document}

\title{{Dark Matter Clusters and Time Correlations  in Direct Detection Experiments}}

\author[a]{Shmuel Nussinov,}
\author[b,c]{Yongchao Zhang}

\affiliation[a]{School of Physics and Astronomy, Tel Aviv University, Tel Aviv 69978, Israel}
\affiliation[b]{Department of Physics and McDonnell Center for the Space Sciences,  Washington University, St. Louis, MO 63130, USA}
\affiliation[c]{Center for High Energy Physics, Peking University, Beijing 100871, China}

  \abstract{Assuming that dark matter (DM) efficiently clusters on various scales we analyse the possible impact on direct {DM} searches. For certain sizes and densities of DM clusters, mutual detector-cluster encounters may occur only once a year or every several years leading to the apparent failure of individual  experiments searching for DM to discover it. If, however, encounters with Earth size and up to $10^4$ times bigger clusters occur about once a year, then  finding time correlations between events in different underground detectors can  lead to DM discovery.
}

\maketitle

\section{Introduction}

The quest for dark matter (DM) comprising {the galactic haloes and about one quarter of the cosmological energy density} has been a holy grail in particle physics research for several decades.
While the mass density of DM in our region of the galaxy was inferred from astrophysical observations to be about $ \rho_{\rm DM} \simeq 0.3 \, {\rm GeV}/{\rm cm}^3$, the mass of the individual DM particles envisioned varies over ninety orders of magnitude from 30 solar mass black holes~\cite{Bird:2016dcv} to ultra-light $10^{-22}$ eV fuzzy DM~\cite{Hui:2016ltb}. Even restricting to particles (rather than meso/macroscoptic nuggets~\cite{Witten:1984rs, Bai:2018dxf, Grabowska:2018lnd, Ge:2019voa} or stars) there are many types of DM~\cite{Feng:2010gw}. These include hot, warm and cold DM, symmetric DM produced thermally or out of equilibrium, asymmetric DM~\cite{Nussinov:1985xr, Kaplan:2009ag},  strongly (self) interacting massive particles~\cite{Spergel:1999mh, Hochberg:2014dra}, dissipative DM, elementary or composite DM, mirror/twin DM~\cite{Berezhiani:2003xm, Foot:2014mia, Berezhiani:1995am, Chacko:2005pe, Chacko:2005un},  
and there are many models of each type.

Some DM particles were independently suggested earlier by efforts to resolve some problems in the standard model (SM), such as the axion for the strong CP problem or the lightest supersymmetric particles in supersymmetric (SUSY) models which addressed the hierarchy problem.  In particular,  DM particles having masses around $\sim$100 GeV and annihilation rates fixed by ordinary weak interactions yield the correct freeze-out relic densities, i.e. the weakly-interacting massive particle (WIMP) miracle, and could be most effectively looked for directly via their coherent interactions with nuclei.
Efforts to directly detect such DM particles, produce them at the Large Hadron Collider (LHC)~\cite{Abercrombie:2015wmb} or find photons, positrons, anti-protons or neutrinos from their annihilation~\cite{Ackermann:2015zua, Aartsen:2012kia} have so far failed. Very stringent upper bounds on the nuclear cross sections of WIMPs have been established by direct detection experiments like PandaX~\cite{Tan:2016zwf, Cui:2017nnn}, XENON1T~\cite{Aprile:2017iyp, Aprile:2018dbl} and LUX~\cite{Akerib:2016vxi}.
These bounds suggested, more than any other evidence, that the classical WIMP paradigm should be abandoned, and a much broader range of lighter DM mass is presently targeted in new experiments~\cite{Essig:2016crl, Budnik:2017sbu, Knapen:2017ekk}.

It has been proposed that in analogy to the standard sector the dark sector consists of many components with mutual interactions at various different scales~\cite{Strassler:2006im}, and that subleading DM components may lead to striking effects~\cite{Randall:2014lxa}. DM may have such minute couplings to the ordinary sector that it will never be directly detected. In stark opposition to this, DM particles could have further properties which would help their discovery if searched appropriately. One particular recent suggestion is that for a certain range of parameters we can get self-destructing subdominant component of DM~\cite{Grossman:2017qzw}. Upon having nuclear interaction, such DM particles release an internal energy which is much higher than the $\sim$ few keV recoil energy. Our present work is very much in the same optimistic vein.

The new features relevant to the discovery of DM discussed here are connected with the clustering of DM.  The ``local'' DM density of $\rho_{\rm DM} \simeq 0.3$ GeV/cm$^3$ inferred from astrophysical observations of its gravitational effects is an {\it average} over a kilo-parsec size region. This average density persists if the DM is not uniform but is clustered in mini-galaxies or smaller micro-haloes. Such structures  naturally arise in the down-up scenarios of cold DM structure formation by gravitationally enhancing primordial density fluctuations~\cite{Goerdt:2006hp}. For concreteness, we will phrase the argument in the context of WIMP searches in the large cryogenic underground experiments although the basic statistical arguments are independent of the type of experiments and apply equally well to searches of axion and other DM particles lighter than conventional WIMPs.

To find the optimal sizes and masses of DM cluster which maximally impact direct DM discovery, we need to use, in addition to the basic astrophysical information on the local DM density of $\rho_{\rm DM}$, various physics parameters of the experiments directively searching for DM. These include the mass $m_{A,Z}$ of the nuclei involved (with $A$ and $Z$ respectively the atomic and proton numbers) or the electron mass $m_e$ for electronic target, the typical linear size $L$ (or the total mass) of the detector, and the duration $t_{\rm expt}$ of the experiment. The effects of clustering may be dynamical, for example the maximization of the recoil energy by matching the target mass and the assumed DM mass $m_{\rm DM}$. However, most of the considerations below will be of simple statistical nature.

The sizes of clusters considered to date vary over a huge range of about 45 orders of magnitude, starting from bound states of size of $1/m_{\rm DM}$ which can be $10^{-20}$ cm to galaxy clusters of ${\cal O} ({\rm Mpc}) \sim 10^{25}$ cm. Clusters in the range of sizes $10^{17}$ cm and Earth masses  $M_\oplus$ (mini-halos) all the way up to size of $10^{25}$ cm and $10^{14}$ solar mass (galaxy clusters) can naturally form by gravitational enhancement of primordial density fluctuations~\cite{Goerdt:2006hp}. Such large clusters have been extensively discussed in the literature, and may impact indirect DM detection by enhancing annihilation rates inside the cluster. We will not consider these here, as we focus mainly on direct detection and on asymmetric non-annihilating DM~\cite{Nussinov:1985xr, Kaplan:2009ag}.\footnote{Constraints from the non-observation of DM at LHC are not affected by the putative DM clustering.  Also constraints based on the non-observation of energetic neutrinos from annihilation of DM in the Sun/Earth are not affected as these annihilations build up over a very long period of DM capture in the Sun/Earth. The same holds true for the recent suggestion to look for traces of DM interactions that happened over astronomically long time in small crystals in stable mineral deposits~\cite{Baum:2018tfw}.}

In Sections~\ref{sec:small} and \ref{sec:intermediate} we address the effect of all clusterings starting with the small sizes and working our way upwards to larger ones.  Before continuing it is worthwhile to point out the methodology and the new results in this paper:
\begin{itemize}
  \item Our discussion does not depend on the detailed design of the direct DM search experiments but only on their final claimed sensitivities. By separating spatially and temporally isolated events with just few keV recoil energy for DM which elastically scatters from nuclei, the background rejection is so good that detecting just few, ${\cal N}_{\rm min}$, ``good'' events during the $t_{\rm expt} \sim {\cal O}$(year) duration of the experiments would strongly suggest a DM origin. (In the following we will often use ${\cal N}_{\min} \approx 6 $). Conversely, not seeing even this small number of candidates for DM collisions imposes upper bounds on the scattering cross sections of DM with nucleons. These are described by the strong exclusion curves in the $\sigma_{\chi N} - m_{\rm DM}$ plane which these experiments keep producing.

      \item  Our working hypothesis, which underlies also the huge experimental effort, is that the discovery of DM in the Xenon experiments  may be just around the corner and the real $m_{\rm DM}$ and $\sigma_{\chi N}$ are  close to the above mentioned excluded region. The latter spans  ``low'' ($\sim 5$ GeV) DM masses with small recoil energies and high multi-TeV masses with reduced DM flux, both resulting in weaker bounds on the DM-nuclear cross section relative to the case of optimal WIMPs of $m_{\rm DM} \sim 100$ GeV with the strongest upper bounds on $\sigma_{\chi N}$. To breach the gap and achieve discovery somewhat bigger experiments and/or longer exposures are then required, or alternatively if clustering happens our suggested search for time correlations may do it.

  \item Using the above and purely statistical considerations, we define the ``critical cluster line'' separating the two regions where efficient cluster formation does or does not affect direct search experiments.
\end{itemize}
The single most important observation is that
\begin{itemize}
  \item If critical clusters in a wide range of sizes between $\sim 10^9$ cm and $10^{14}$ cm exist and contain a sizable fraction of DM, then just comparing the times of the handful of ``relatively good events'' collected in the present large underground experiments located in three different continents  could almost immediately lead to DM discovery.
\end{itemize}
In Section~\ref{sec:large} we will discuss the main new result of our work: the remarkable opportunity that certain clustering offers for DM discovery by looking for unexpected time correlations between the events recorded in the underground facilities in different continents. The clusters required in this case are very special and very different from the conventional gravitational clustering. We discuss their stability against tidal disruption in Section~\ref{sec:stability} and address the issue of their formation in Section~\ref{sec:formation}, before concluding in Section~\ref{sec:conclusion}. We also comment on the narrower recoil energy distribution arising when the DM is clustered, on the likelihood that the small clusters/blobs of interest  will reach the underground detectors, and on the effective cooling of the WIMPs via coalescence into blobs. 

\section{Smaller clusters}
\label{sec:small}

We start with the case of small clusters, defined by having sizes $R$ smaller than the typical linear size $L \sim {\cal O}({\rm meter})$ of present DM detectors. While the composite DM and blobs/grains of DM belonging in this category have been considered by many authors, we still briefly review these for completeness. This much simpler case helps illustrate the important feature which defines for all clusters  the regions where clustering has important effects.  Also some points  may have been rather poorly emphasized before or even be actually novel.

To simplify the argument, we assume throughout spherical clusters of radius $R$ with uniform DM number or mass density inside the cluster, which is enhanced by a factor $E$ relative to the average $\rho_{\rm DM}$. We further assume the same radius and enhancement factor for all clusters and that  a significant portion of DM is inside them.

A common  important feature of all clusters is that they form prior to the formation of the larger  structures and in particular the galaxies. This is quite obvious for the case of the smaller clusters/blobs that we discuss here as these are governed by short range forces which are much stronger than gravity. Also the galactic virial velocities of order $v_{\rm Virial}^{\rm (galaxy)} \sim v_{\rm Virial} \sim 300 \, {\rm km}/{\rm sec}$ are far larger than the small escape velocity from the dilute clusters of main interest. This then will impede their formation at late $z < 10$ epoches. Assuming that the clusters formed before the galaxy, the gravitational field of the galaxy and halo will impart to them the same velocity spectrum peaked around $ v_{\rm Virial} \sim 300$ km/sec with an approximately isotropic distribution in the halo rest frame, as it would to individual unclustered DM particles.
How and when then can clustering affect direct detection?

The smallest ``clusters''  are composite DM, i.e. stable bound state of elementary particles. The dynamical question of how  the cross section of this composite DM for scattering with ordinary SM particles differ from the sum of cross sections of the constituents is of some interest and has been discussed in some details~\cite{Krnjaic:2014xza, Hardy:2014mqa, Coskuner:2018are}. Form factor of loosely bound extended composite DM will decrease the net cross section~\cite{Gelmini:2002ez} whereas coherence effects, analog of the $A^2$ factor in spin-independent WIMP-nuclei scattering, will increase the elastic scattering when the composite's size  is smaller or equal to that of the nuclear target. Since the kinetic energy of the composites is negligible as compared with their binding, we will view them as effectively elementary WIMPs.

\begin{figure}[tb]
\centering
  \begin{subfigure}[b]{0.18\textwidth}
  \includegraphics[width=\textwidth]{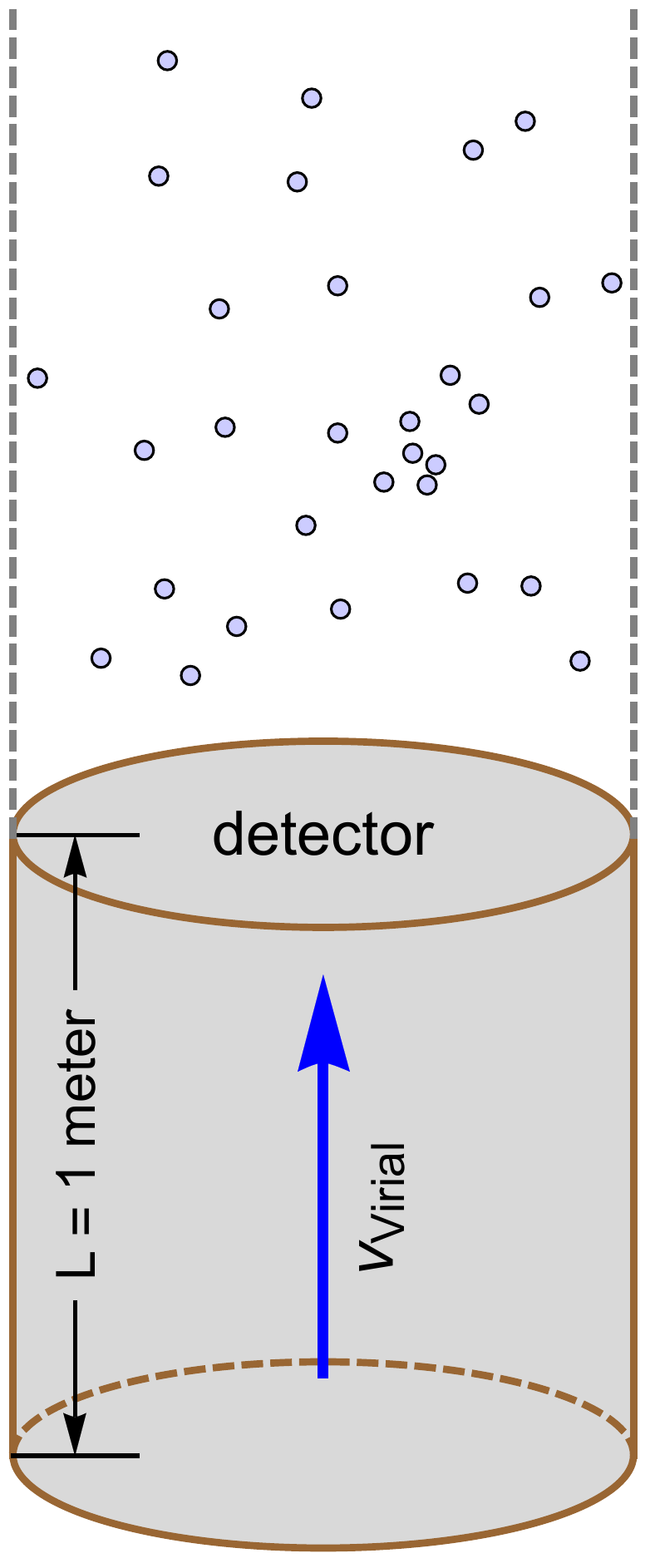}
  \caption{}
  \label{fig:1a}
  \end{subfigure} 
  \begin{subfigure}[b]{0.18\textwidth}
  \includegraphics[width=\textwidth]{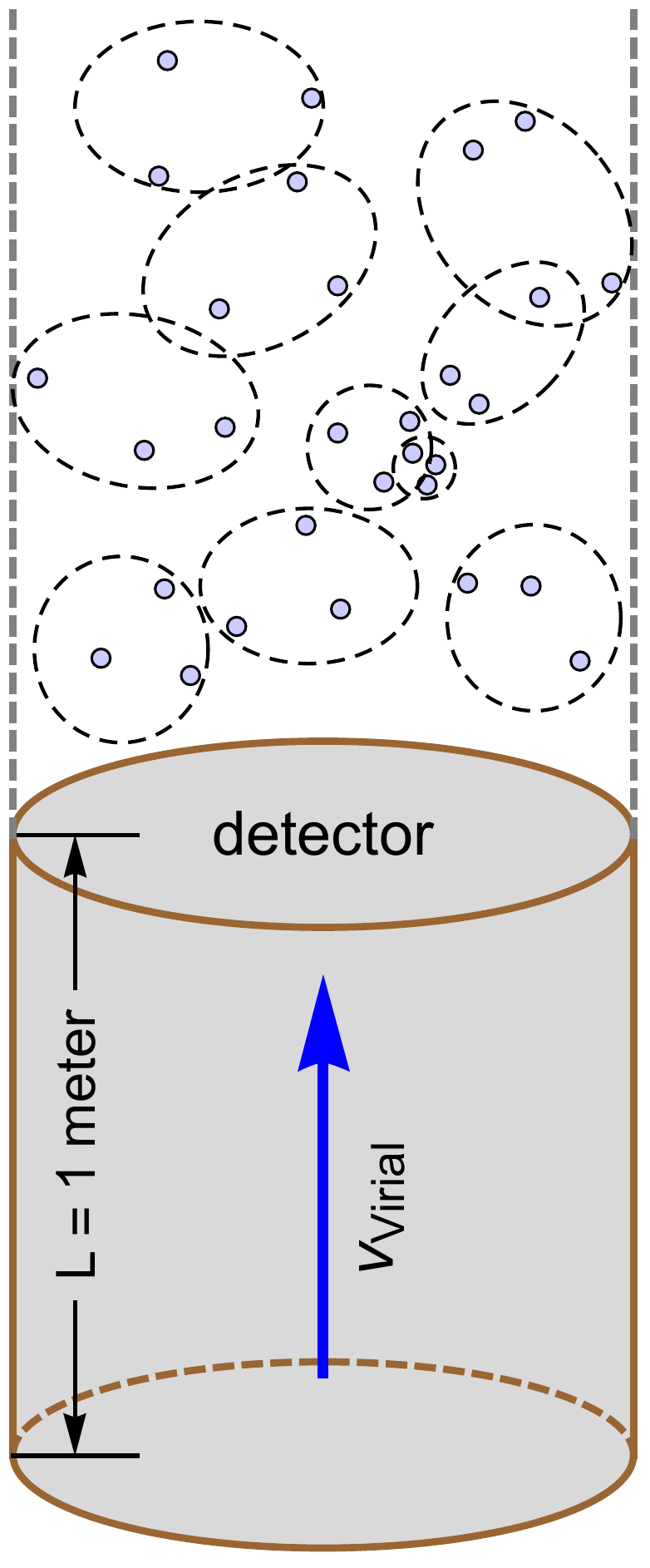}
  \caption{}
  \label{fig:1b}
  \end{subfigure} 
  \begin{subfigure}[b]{0.18\textwidth}
  \includegraphics[width=\textwidth]{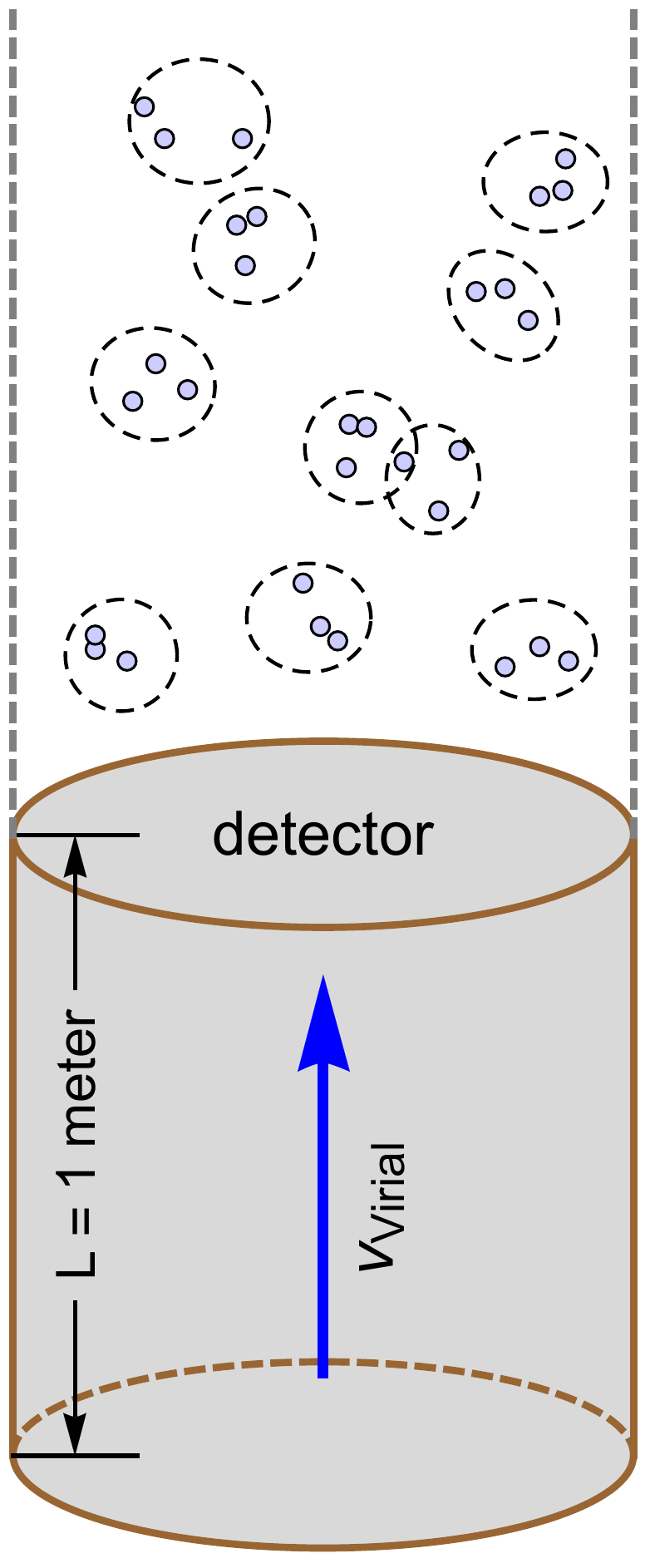}
  \caption{}
  \label{fig:1c}
  \end{subfigure}
  \begin{subfigure}[b]{0.12\textwidth}
  \centering
  $\cdots\cdots$
  \caption{}
  \label{fig:1d}
  \end{subfigure}
  \begin{subfigure}[b]{0.22\textwidth}
  \includegraphics[width=\textwidth]{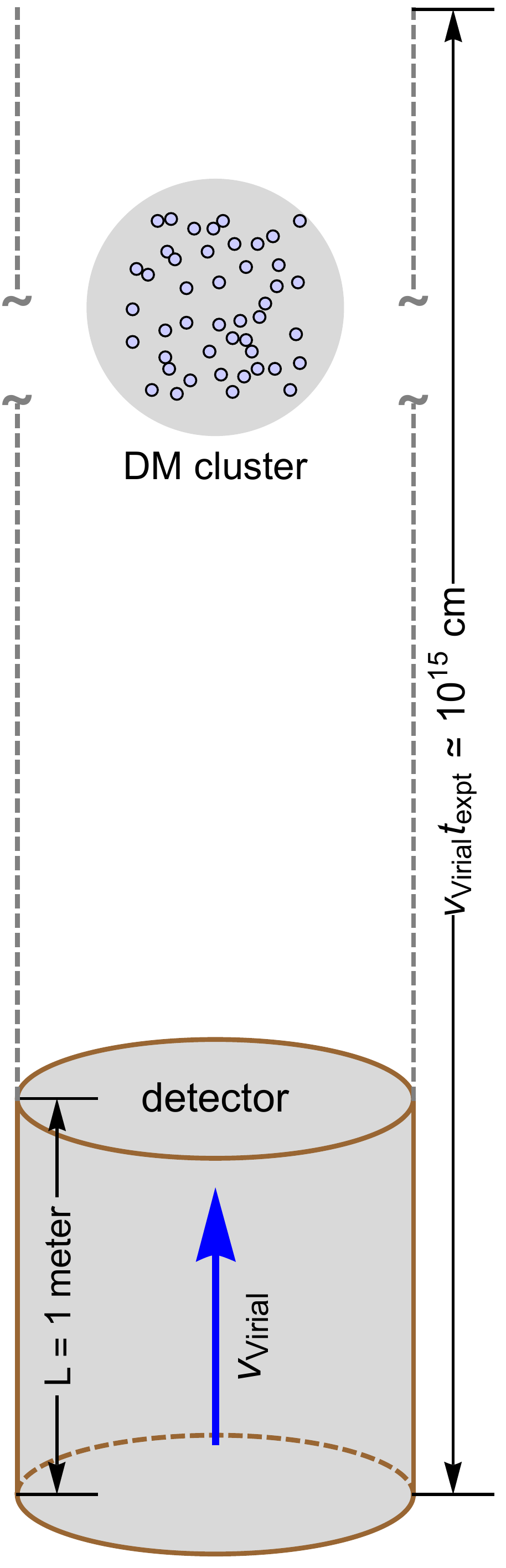}
  \caption{}
  \label{fig:1e}
  \end{subfigure}
  \caption{(a) Unclustered individual DM particles collected by the detector of linear size $L = 1$ meter. We envision that all DM particles are at rest and only the detector moves up with a relative velocity indicated by the blue arrow of $v_{\rm virial} \simeq 300$ km/sec. In reality the DM particles move as well, but since only their velocity relative to the detector is relevant, this description is adequate. (b) We indicate by circles the triplets of DM particles which form small clusters, and (c) shows the actual clusters of DM triplets. The horizontal dots in (d) describe the continuation of this process where larger and larger clusters are being made. Finally (e) shows the limiting case where all DM particles which would have been collected in a year form one grain of the critical mass in Eq.~(\ref{eqn:criticalmass}). We note that until the last stage (e) is reached there is no difference in the number of expected nuclear interactions and of the timings of these interactions  between the case of clustered and unclustered DM. }
  \label{fig1}
\end{figure}

When the DM number density in the clusters is less than nuclear density, the DM particles therein scatter incoherently and independently from nuclei in the detectors. This is most clearly the case for the dilute large clusters of main interest. Also the small DM - ordinary matter scattering cross sections exclude any shadowing effects by the different WIMPs in a grain. This along with the identical velocity spectra of clustered and unclustered DM suggest that over a very large range  formation of both large and small clusters will have no effect on the direct DM searches.  Indeed if the DM particle hit our detector singly as in Fig.~\ref{fig:1a}, or in groups of two or three particles, as in Fig.~\ref{fig:1b}, and \ref{fig:1c}, etc, there will be no change in the number or nature of DM-nuclear collisions, neither in their expected uniform distribution over the running time $t_{\rm expt}$ of the experiment.

But how far can  this continue?

\begin{figure}[tb]
  \centering
  \includegraphics[height=0.55\textwidth]{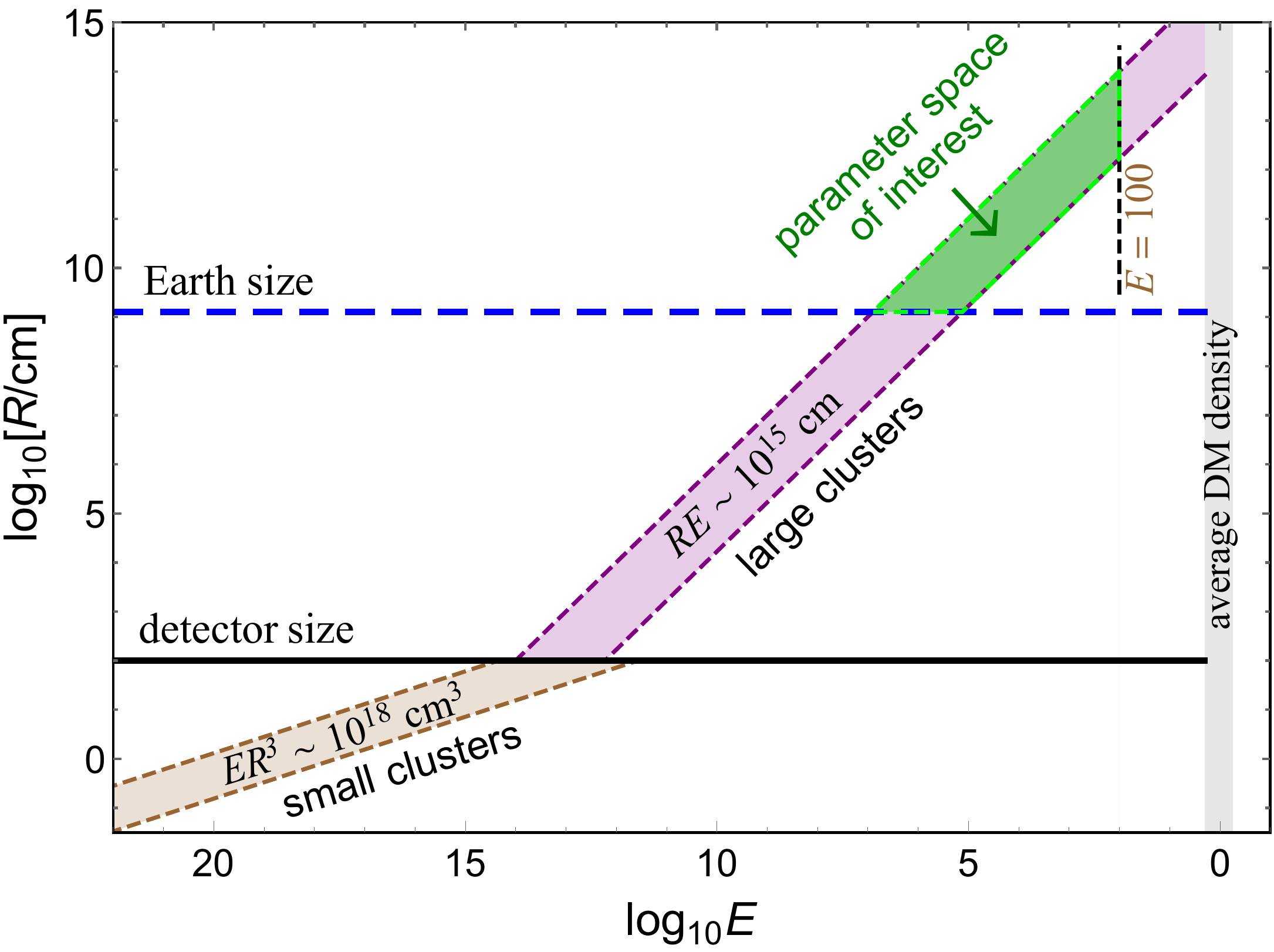}
  \vspace{-5pt}
  \caption{The critical region for small clusters defined by $ER^3 \simeq 10^{18\pm 1} \, {\rm cm}^3$ is indicated by the brown diagonal strip in the left-lower part of the figure. The critical region for large clusters defined by $ER \simeq 10^{15\pm1}$ cm is indicated by the purple diagonal strip in the right-upper part of the figure.
  The significance of the shaded areas is -- as explained in the text -- that for clusters with $R$ and $E$ parameters lying therein, the probability that the detector and cluster will meet during one year is of order one. This further ensures that in each year we will have several -- ${\cal N} \sim 6$ -- nuclear collision events occurring during the cluster detector overlap.
  The detector and Earth sizes are also indicated by the horizontal solid black and dashed blue lines.   The green part of the diagonal stripe is the region of prime interest where all the detectors on Earth {\it jointly} encounter the same DM cloud. This leads to {\it coincident events} during a time window of $(30 - 3 \times 10^5)$ sec.
   }
  \label{fig:cluster}
\end{figure}

The boundary separating the region in the $R - E$ plane where clustering does or does not affect the DM detection experiment is fixed in all cases by the same simple statistical consideration. It corresponds to the case where during the typical running time of $t_{\rm expt}$ of the experiment only one or few encounters with a grain (or cloud) happens, as shown in Fig.~\ref{fig:1e}. The mass of that critical DM grain is then the total DM mass flux during a running time $t_{\rm expt}$ through the detector of area $A$:
\begin{equation}
\label{eqn:criticalmass}
M_{\text{critical}}  = \rho_{\rm DM} \, v_{\rm Virial} \, A \, t_{\rm expt} \end{equation}
which is independent of the mass $m_{\rm DM}$ of the individual DM particles. Using representative numbers from the large liquid Xenon experiments with $t_{\rm expt} \sim {\cal O} ({\rm year})$ and $A \simeq L^2 \simeq 10^4 \, {\rm cm}^2$, we find $M_{\text{critical}} \simeq 3 \times 10^{18}$ GeV. This corresponds to $R^3E \simeq 2 \times 10^{18} \, {\rm cm}^3$ (see brown strip in Fig.~\ref{fig:cluster}) and a total number of DM particles traversing the detector
\begin{equation}
\label{eqn:Ntotal}
{\cal N}_{\rm total} \ = \ 3 \times 10^{18} \times \ \left( \frac{m_{\rm DM}}{\rm GeV} \right)^{-1} \,.
\end{equation}
Grains  much heavier than the above  critical mass require far larger detector. An extreme example is provided by the strange quark or other nuggets~\cite{Witten:1984rs, Bai:2018dxf, Grabowska:2018lnd, Ge:2019voa} where encounters with the whole Earth may generate observable seismic effects~\cite{DeRujula:1984axn}.

The events generated by critical or larger grains hitting the detector are very different from those due to single DM interactions in the unclustered case. All the ${\cal N}_{\rm event} \sim {\cal N}_{\rm min}$ interactions induced by the one per year critical grain hit will now happen while the grain passes through the detector during $L/v_{\rm Virial} \simeq 3 \times 10^{-6} \, {\rm sec}$ which is a tiny fraction $( \simeq 10^ {-13})$ of the year. Also the interactions of DM particles within the grain with nuclei in the detector will be restricted to a cylinder along the direction of the incoming grain, which is defined once the event number ${\cal N}_{\rm event} \ge 2$. The effective volume of this cylinder $\pi R_{\rm eff}^2 \times L$ is only a small fraction $f \simeq \pi R_{\rm eff}^2/L^2$ of the detector's volume. Here $R_{\rm eff}$ denotes either the radius of the grain or, when the grain is smaller than the spatial resolution limit of the detector, $R_{\rm eff}$ is the size of the latter resolution limit. For $R_{\rm eff}=1$ cm and $L= 100$ cm,
$f \simeq 3 \times 10 ^{-4}$ is quite small.  Furthermore, if $t_i$ are
the times of consecutive interactions at locations $\vec{r}_i$ inside the detector, then all {$v_i = |\vec{r}_i-\vec{r}_{i+1}|/{|t_i-t_{i+1}|}$} should have a common value $v_i=v $ which is the velocity of the grain. The facts that all the hits are located on a single line or within a thin cylinder, that the common velocity inferred is about $v \simeq v_{\rm Virial} \simeq 300 \, {\rm km}/{\rm sec}$, and that this velocity tends to  be in  the direction of the ``WIMP wind'', help exclude background events due to neutrons or relativistic muons, and pin down the DM source of the signal.

Some searches for massive milli-charged particles causing multiple ionizations along their straight line paths have been made~\cite{Hall} and a discovery of such particles would be equally exciting. Despite some superficial similarities, the nature of the individual low energy nuclear recoil events in the case of the DM grain would clearly distinguish between the two different cases.

\section{Large size clusters}
\label{sec:intermediate}

The main difficulty of detecting DM which is inside small grains/blobs of critical mass is that all the nuclear interactions occur within the very short time while the grain traverses the detector. Without using all the further  information described in the previous section, such an event would be discarded as some background noise. If the blobs are somewhat heavier than critical, at most one of the three large underground detectors presently involved in direct search experiments PandaX, XENON1T and LUX will encounter such a grain in its ${\cal O} ({\rm year})$ running time.  Any claims of a discovery  will then be clearly disputed by the other two experimental groups.

The difficulties will be further exacerbated if the clusters are not tiny grains but of size $R$ which approaches the $L \sim 1$ meter size of the detector. In this case the nuclear recoils  will tend to be uniformly spread over the volume of the detector and the exclusion of background by the collinearity of the recoils is less efficient.

\begin{figure}[tb]
  \centering
  \includegraphics[height=0.58\textwidth]{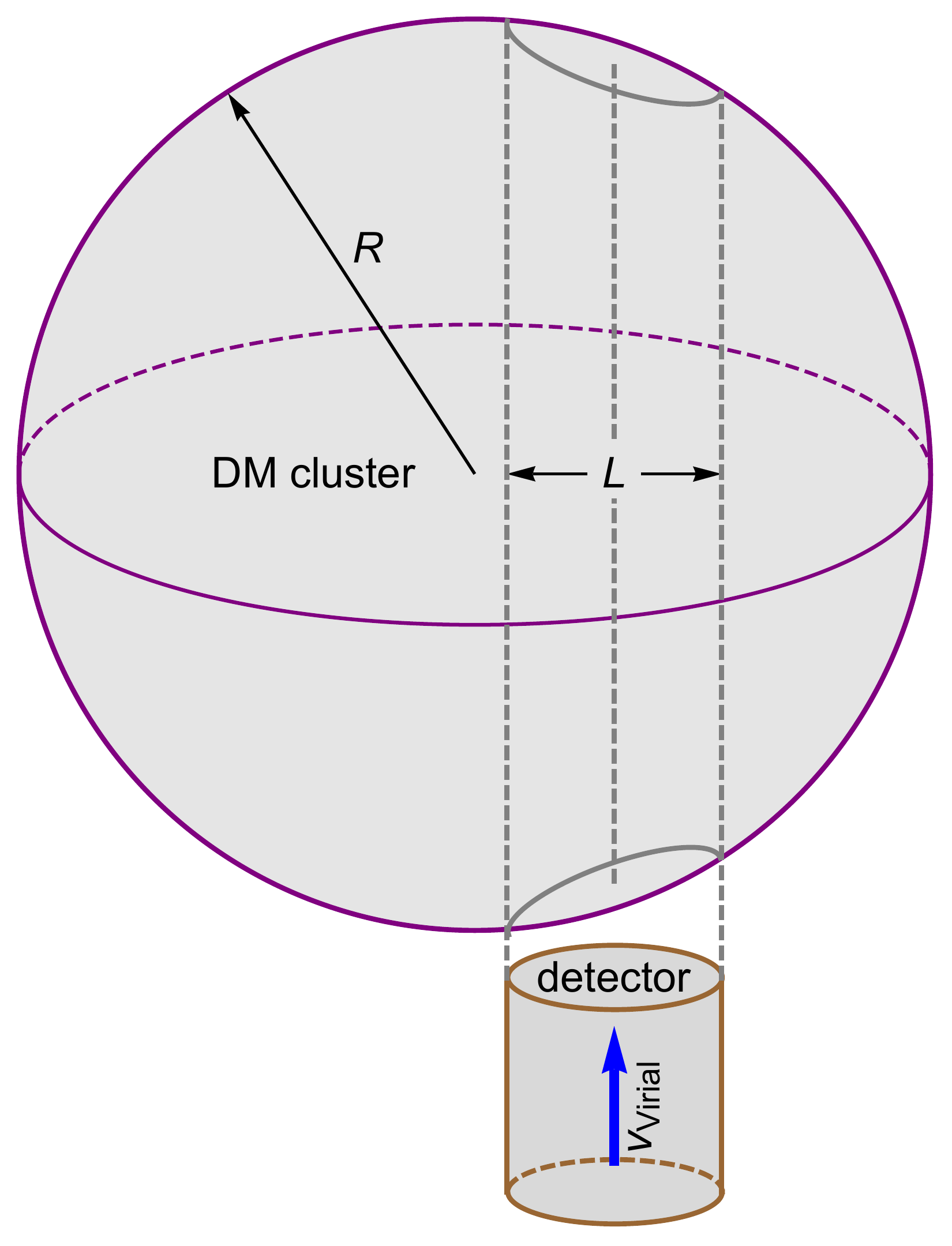}
  \vspace{-5pt}
  \caption{The dashed lines indicate the cylindrical portion of the large cluster of radius $R$ which a detector of size $L < R$ traverses. We refer to this as the cylindrical overlap region of volume $\pi L^2 R$. }
  \vspace{-8pt}
  \label{fig:intermediate}
\end{figure}

We next proceed to clusters which are larger than the detector $L$. A key difference relative to the case of small ($R<L$) clusters is that the border (critical) line of relevance in the present case is no longer defined by the requirement that the clusters have the critical mass in Eq.~(\ref{eqn:criticalmass}). The point is that only the WIMPs within the cylindrical region of volume $\sim \pi L^2R$ shown in Fig~\ref{fig:intermediate} which is traversed  by the detector within the larger DM cluster/cloud can interact with the nuclei inside the detector. Since the density within the cloud is $E\rho_{\rm DM}$, the total WIMP mass within this region is
\begin{equation}
 M_{\rm cylindrical} \ \simeq  \ \pi E\rho_{\rm DM} L^2 R \,.
\end{equation}
Demanding that this mass be $M_{\rm critical}$ of Eq.~(\ref{eqn:criticalmass})  so as again to allow ${\cal N}_{\rm event} \sim {\cal N}_{\rm min}$ nuclear collision per detector-cluster encounter, we find the new condition:
\begin{equation}
\label{eqn:ER}
ER = v_{\rm Virial} t_{\rm expt} \ \simeq \ 10^{15} \, {\rm cm} \,,
\end{equation}
which is indicated by the purple strip in Fig.~\ref{fig:cluster}. Here we have taken $t_{\rm expt} \simeq 1$ year. We note that since the cross sectional area  $\pi R^2$ of the cloud is $(R/L)^2$ times larger than that of the detector the probability of our detector encountering any give cloud is $(R/L)^2$ times larger than that of the critical grain hitting the detector in the previous case of small clusters. However, the total volume of a new larger cloud is $(R/L)^2$ times larger than that of the cylindrical overlap region and hence is $(R/L)^2$ times more massive than that of the ``overlap region'' which, by definition, has the same mass as the critical grain. Since in both scenarios the grains or the clouds have to contain the same total mass, the number density of the clouds should be $(L/R)^2$ smaller than that of the critical grains in the previous case. The net effect of the enlarged areas and reduced number of clouds is that we have the same rate of new cloud-detector encounters as the rate of critical grain-detector encounters in the previous case.  The condition in Eq.~(\ref{eqn:ER}) will then indeed correspond to having ${\cal N}_{\rm event} \sim {\cal N}_{\rm min}$ nuclear collision per year in the detector in the case
of large ($R>L$) clusters.

Eq~(\ref{eqn:ER}) and the encompassing discussion are of utmost importance and underlie the main result of our paper. We therefore add here further two lines of argument which lead to it:
\begin{itemize}
  \item Assume that all the $n_{\rm DM} D^3$  DM particles within any one of non-overlapping, neighboring cubes of side $D$ eventually form our cluster of size $R$ and the same number of DM particles: $(4\pi/3) R^3 E n_{\rm DM}$.  The spacing between clusters then is $D \sim [4\pi/3 E]^{-1/3}R$ yielding a density of clusters of $n_{\rm clusters} = 1/D^3= (3/4\pi) (R^3E)^{-1}$.   Since the cross section of collision with a cluster is $\sigma_{cluster} \sim \pi R^2$ we find for the mean free path (MFP) -- essentially the distance traveled between collisions is
      \begin{eqnarray*}
      l_{\rm MFP} = \frac{1}{n_{\rm cluster}\sigma_{\rm cluster}} \sim RE
      \end{eqnarray*}
      If we further take $RE \sim 10^{15}$ cm, then assuming that the (relative) velocity is $v \sim v_{\rm Virial} = 300$ km/sec the time spent between successive collisions $l_{\rm MFP}/v_{\rm Virial} = 3 \times 10^7$ sec is indeed the desired year benchmark.
  \item Since the density of DM particles within the clusters is $E$ times higher then in the original unclustered case, the clusters occupy only $1/E$ of the space so as to keep the same total number of DM particles. Any object -- say the DM detector -- which is moving relative to the clusters in some random direction, should therefore spend only $1/E$ of the time inside clusters. Since it takes $\Delta t =R/v_{\rm Virial}$ to traverse a cluster this implies that the time intervals  between encounters with clusters should be $E \Delta t =ER/{v_{\rm Virial}}$ which by our choice of $ER=10^{15}$ cm is about a year.
\end{itemize}

To further clarify Fig.~\ref{fig:cluster}, in particular the ``large cluster'' part, let us briefly reiterate the underlying rational. In the unclustered case, the total number of DM particles traversing our detector during a year-long run is  $\pi n_{\rm DM} L^2 (v_{\rm virial} T)$ with $T = $ year. The last term in the parenthesis -- the total distance traveled -- is $10^{15}$ cm. The experimental groups state that when the underlying micro-physics parameters $\sigma_{\chi n}$ and $m_{\rm DM}$ lie on the famous exclusion curves, this ensures some minimal number ${\cal N}_{\rm min}$ of order a few, $\simeq 6$,  DM-nuclear collisions. Our critical line is defined by $RE \simeq 10^{15}$ cm. As we show in detail, this is the average  distance travelled between consecutive encounters with two large clusters. Requiring that $RE = v_{\rm Virial} T$ with $T = $ year, we will have such an encounter occur once a year. As the detectors transverse the clump of size $R$, a total number $\pi (E n_{\rm DM}) L^2 R$ of DM particles is encountered. This is equal to the total number $\pi n_{\rm DM} L^2 (v_{\rm virial} T)$ with $v_{\rm virial} T = RE$ encountered in the unclustered case above. This in turn guarantees the same number ${\cal N}_{\rm min}$ of nuclear collisions. 

 In general clustering in the regions above the critical lines tends in
both cases of small and large clusters to decrease the prospect that any single direct detection experiment will discover DM. Indeed, unless the size of the underground detectors dramatically grows, which seems rather unlikely, the blob/cloud encounters with the detector will become so rare that even running over many years will not suffice to benefit from the higher concentration within these clusters.  As an example, the $E - R$ regions corresponding to the ordinary gravitationally formed clusters lie well above the $ER= 10^{15}$ cm line. The chance that we are within any such cluster are very small and the  relatively dry spells when we are somewhere between them can last for many years. Conversely, we have seen that when we are under the critical lines the clustering has no effect on DM detection, except for the possible dynamics of the interactions in the case of DM composites which are treated as a new type of DM.


There is still one feature connected with the spectrum of recoil energies which is of interest. Because the tiny escape velocities from the DM clouds are much smaller than the velocity of the cloud (the latter being of order $v_{\rm virial} \sim 300$ km/sec), all the DM particles within any given cloud share the same velocity and move in the same direction. While the large Xenon detectors supply no directional information,  the measured distribution of recoil energy will be affected. For  unclustered DM this distribution reflects both the initial  DM velocity distribution, usually taken to be a broad shifted Gaussian, and the fraction of energy transmitted to the nuclear recoil which depends on the mass ratio of the DM and nuclear target and the scattering angle (whose distribution is determined by the nuclear form factor). Because all DM particles in the cloud share the same (albeit unknown)  velocity/kinetic energy,  we expect the width of the resulting recoil energy distribution to be significantly smaller.  This
feature is common to  {\it all} the clusters considered.

As we proceed to larger and larger critical clusters, only the times of traversing the cloud change. These times grow from 3 micro-seconds for $R= L=1$ meter to $\delta t \simeq 1$ year for the largest clouds of size $R=10 ^{15}$ cm with the minimal density enhancement $E=1$ where we have in effect
returned to the original unclustered case.  Thus it appears that
formation of critical clusters over this very large span of linear
sizes will not enhance the prospect of DM discovery in any single
direct search experiment.\footnote{A possible exception would arise if distributed set-ups were used with say two or more parts of the detector deployed underground at some distance $d \gg L$  where coincident counts might arise if $R > d$. While this is clearly possible for small solid state
set-ups such as CDMS, the novel multi-tone Xenon experiments are much
more efficient and cheaper to build and maintain as one large unit.}

\section{Time correlations between different DM experiments}
\label{sec:large}

The key observation which we make in this paper is that the above
discouraging conclusion is avoided if we compare the
timings of the various events in the different underground experiments. Of particular interest are the clusters whose size falls in the
range:
\begin{eqnarray}
\begin{aligned}
\label{eqn:Rrange}
R \ \simeq \
(10^9 - 10^{13}) \, {\rm cm}
\ {\simeq} \ (1- 10^4) \, R_{\oplus} \,,
\end{aligned}
\end{eqnarray}
with $R_{\oplus}$ the Earth radius. This is shown in green in Fig.~\ref{fig:cluster}, with an enhancement factor $E >100$. The ${\cal N}_{\min}$ ``quota'' of scattering events
expected to occur during a year in each of the ${\cal N}_{\rm det}$
terrestrial detectors should then all occur during the encounter with a
single large cluster which simultaneously overlaps all the detectors.
The duration of this encounter
\begin{eqnarray}
\delta t \ \simeq \
R/v_{\rm Virial} \ \simeq \
(30 - 3\times10^5) \, {\rm sec} \,,
\end{eqnarray}
is the length of the common time interval during which all these ${\cal N}_{\rm det} {\cal N}_{\min}$ events should occur. Using one interaction in one of the detectors as our reference point in time, the probability $P$ that all other events will occur within ($10^{-6}-10^{-2}$) of a year near this reference time rather than be uniformly distributed over the year is
\begin{eqnarray}
\begin{aligned}
\label{probability}
P \ = \ 10^{-6({\cal N}_{\rm det} {\cal N}_{\rm min}-1)} \quad {\rm to} \quad 10^{-2({\cal N}_{\rm det} N_{\rm min}-1)} \,.
\end{aligned}
\end{eqnarray}
For ${\cal N}_{\rm min} = 6$ and just two jointly operating direct DM
detectors (${\cal N}_{\rm det} =2$), $P \simeq (10^{-66}-10^{-22})$ is incredibly small. The extreme assumption that ${\it all}$ DM is segregated within these clusters is
not necessary. If only $1/3$ of DM is within the clusters considered and the rest remaining unclustered (in this case the limits from the standard analysis still apply), then only
${\cal N}_{\rm det} {\cal N}_{\rm min}/3$ events would be expected to
occur within the above time window. This increases the probability for coincidences to $P \simeq (10^{-18}- 10^{-6})$, which is still very small.\footnote{Enhancing $RE$ by a factor of $k >1$ prolongs the running time required to meet a cluster to $k$ years but causes a $k$ fold increase of the number of interactions expected during the encounter with the DM cluster, which, in turn, decreases the probability of random coincidence from the $P$ in Eq.~(\ref{probability}) to $P^k$. Furthermore, directional information is not available for large clusters where the DM interactions in any given detector are uniformly distributed over the volume of the detector. Still for Earth size and somewhat smaller clusters, the time ordering and time separation of the groups of events in different detectors reflect to some extent on the magnitude and direction of the cluster's velocity.} As noted above the common velocity of all the DM particles in the cluster implies that all the collisions in a single detector-cluster encounter will have the same maximal cutoff energy, which is another significant indicator helping beat backgrounds even when the fraction of clustered DM particles is small. This clearly applies also to the small cluster case above.

Noises in the widely separated independent experiments are not
correlated and the coincident events will be highly significant. Finding coincidence between the detectors in {the INFN Laboratori Nazionali del Gran Sasso}, the Jinping Underground Laboratory and the Sanford Underground Laboratory located in three different continents
requires collaborations between the different teams. Such collaborations were extremely successful in the multi-messenger investigations of rare astronomical events, like the recent two neutron star merger seen in gravitational waves and in the electromagnetic spectrum. Since there is no need to alert any of the experimental groups as they are continuously ``observing'', the effort required is minimal. All we need is to compare, {\it after} the completion of experiments which run in parallel, the recorded times of the handful of ``very good'' and also of the ``just good'' events seen in different detectors.

In the effort to find the $\sim {\cal N}_{\min}$ splendidly isolated, genuine WIMP-nuclei collision events with optimal signatures in, say, both ionisation and scintillation signals, the experimental groups often discard many events which are not as clean. Nearby times of such ``background'' events in the different detectors would elevate them to be part of the true signal.


We note that even barring the required minimal collaboration between the  strongly competing experimental groups  some preliminary indications of the existence of optimal clustering can already be seen by each group separately. These will manifest by having a subset  say two or three of the events occurring with time intervals which are sufficiently shorter that the ${\cal O} ({\rm year})$ running time. If these time patterns (and the tendency of the events in these time intervals to have more similar energies than some random pairs or triplets)  are statistically significant, then time intervals which are longer than 30 sec may correspond to the passage of the optimal clusters which we focus on. Once some groups find such time clustering they can compare these times, and if these happen to be close then the exciting possibility of a true DM discovery should be entertained.

\section{Cluster stability}
\label{sec:stability}

\subsection{Stability of small clusters/grains}

In the case of small clusters/grains, we clearly do not have to worry about the tidal effect. However, we worry about the effect of the transit through Earth. If the DM in any clusters/grains undergoes ${\cal N}_{\rm min} \sim 6$ collisions with the Xenon nuclei in the $L \sim 1$ meter size detectors, it can undergo $N_{\rm collision} \sim 10^6 -10^7$ times more collisions while traversing the Earth enroute to the detector which is $\sim 2$ km underground.\footnote{DM particles coming from the upper hemisphere (as viewed from  each detector) travel in Earth at most 160 km and $\sim 87$\% of all WIMPs traverse at most $\sqrt2 R$ along paths which lie mainly in the mantle of density similar to liquid Xenon but with typically nuclear $A$ values which are only $\sim 1/5$ of $A \simeq 130$ of Xenon. For the spin-independent interactions of main interest this reduces the expected number of such collisions by a factor of 25. Also assuming $m_{\rm DM} > m_{\rm Xenon}$ (with $m_{\rm Xenon}$ the Xenon nuclei mass) the energy transferred to the DM particle in the grain is only $1/5$ of the initial kinetic energy $m_{A} v_{\rm Virial}^2$ of the nuclei in the grain's rest frame of interest here.}  For these compact, relatively tightly bound critical blobs/small grains, we need to compare the total internal energy $W$ transferred to the grain and the total binding energy of the grain $B= {\cal N}_{total} \epsilon$, with ${\cal N}_{\rm total}$ being the total number of DM particles in the critical grain given in Eq.~(\ref{eqn:Ntotal}) and $\epsilon$ the binding energy inside the grain per DM constituent. With the energy transferred in each collision being = $\frac12 m_{A,Z} v_{\rm Virial}^2 \sim 10$ keV, we have $W \sim 10^{10}$ eV for $10^6$ collisions. Hence once $\epsilon$ is larger than the tiny value of $10^{-2}$ eV, and $m_{\rm DM} \leq 10^6$ GeV so that ${\cal N}_{\rm DM} > 10^{11}$, we have $B > W$ and no appreciable disintegration of the blob is expected.

\subsection{Stability of large clusters/clouds}

We start this section by explaining how, counter to simple intuition, the smaller clusters  with tiny escape velocities are more stable than the large clusters/sub-halos with much larger escape velocities.
This is precisely because the large ratio of the physical sizes! Consider an external  mass $M_{\rm ext}$ at a distance $L$ from a small cluster with radius $r$, mass $m$ and escape velocity $v_{\rm escape}^2= G_N m/r \propto \rho_{\rm small} r^2$ with $\rho_{\rm small}$ the DM density inside the small clusters, or a large cluster with radius $R$, mass $M$ and escape velocity $V_{\rm escape}^2=G_N M/R \propto \rho_{\rm large} R^2$ with $\rho_{\rm large}$ the DM density inside the large clusters. We assume that $r \ll R < L$.
The tidal accelerations operating (for a common time $\delta t = L/v$ with $v \sim 300 \, {\rm km/sec}$) cause a velocity change of DM particles  inside the clusters of
\begin{eqnarray}
\delta v \ = \ a_{\rm tidal}^{\rm (small)} \delta t
\ = \ \frac{G_N M_{\rm ext} r}{L^3} \frac{L}{v}
\ = \ \frac{G_N M_{\rm ext}r}{L^2v}
\end{eqnarray}
in  the small clusters and
\begin{eqnarray}
\Delta v \ = \
a_{\rm tidal}^{\rm (large)} \delta t
\ = \ \frac{M_{\rm ext} R}{L^3} \frac{L}{v}
\ = \ \frac{G_N M_{\rm ext} R}{L^2v}
\end{eqnarray}
for the large clusters.  The fractional changes of velocities are then:
\begin{eqnarray}
\label{eqn:delta}
\delta \ = \ \frac{\delta v}{v_{\rm escape}}
\ = \ \frac{\sqrt{3G_N}M_{\rm ext}}{\sqrt{4\pi \rho_{\rm small}} L^2 v}
\end{eqnarray}
for the small cluster, and
\begin{eqnarray}
\label{eqn:Delta}
\Delta \ = \ \frac{\Delta v}{V_{\rm escape}}
\ = \ \frac{\sqrt{3G_N}M_{\rm ext}}{\sqrt{4\pi \rho_{\rm large}} L^2 v}
\end{eqnarray}
for the large clustes/sub-halos.

Density profiles of the Milky Way and those of eight of its large sub-halos are presented in Fig.~4 of~\cite{Diemand:2009bm}. The sub-halo densities are smaller than the galactic densities by factors of $3 - 30$, and the densities of our conjectured clusters are $10^2 - 10^6$ times the local Milky Way halo density $\rho_{\rm local}$.
Also a recent paper by Arvanitaki {\it et al}~\cite{Arvanitaki:2019rax}
presents in Fig.~11 the densities of Milky Way sub-halos expected in the
standard cold DM scenario. These densities (the values of scale density $\rho_s$ in assumed NFW profiles which are roughly the average densities) are decreasing with the halo masses, and for $M \sim (10^9- 10^{10}) \, M_{\odot}$ are smaller than $2 \times 10^{-3} M_\odot/{{\rm pc}^3}$
and therefore much smaller than the densities of the clusters considered here. Consequently Eqs.~(\ref{eqn:delta}) and (\ref{eqn:Delta}) imply that $\delta \ll \Delta$, implying that the relatively small clusters of interest are more immune to tidal distortions as claimed.


Before embarking on a more detailed discussion of  tidal disruption  we would like to broaden the discussion in the following way. Unlike in an earlier version of this work,  we will not restrict it to tidal disruption and in the next section to formation of the ``large clusters'' in the context of gravity alone. Rather we will allow for an additional attractive DM-DM force of a range somewhat larger than the size of the clusters of interest. Effectively this enhances by $G_N \to f G_N$ ($G_N$ is the gravitational constant) with a positive $f>1$ the ``internal'' gravity inside the clusters. It does not however change the purely gravitational interactions of the cluster and the DM particles in it with external baryonic objects. The gravity-like acceleration at the surface of the cluster is:
\begin{eqnarray}
\label{eqn:gcls}
g_{\rm cls} \ = \
\frac{f G_N M_{\rm cls}}{R^2} \ \simeq \
\frac{4\pi}{3} f G_N \rho_{\rm DM} ER \ \simeq \
1.5 \times 10^{-16} \times f {\rm cm}/{\rm sec^2}
\end{eqnarray}
where $M_{\rm cls}$ is the cluster mass. The escape velocity spans the range
\begin{eqnarray}
v_{\rm esc}^2 \ = \
g_{\rm cls}R \simeq
(10^{-7} - 10^{-3}) f \, ({\rm cm}/{\rm sec})^2
\end{eqnarray}
for the clusters of interest with radii $10^{9} \, {\rm cm} < R < 10^{13}$ cm.

The galactic tidal acceleration
\begin{equation}
a_{\rm tidal} \ = \ \frac{G_N M R}{l^3}
\end{equation}
with $M \sim 2\times 10^{11} M_\odot$ the total mass and $l \sim 7$ kpc, is smaller by about a factor of 10 than the tiny surface gravity in Eq.~(\ref{eqn:gcls}) of the large optimal clusters and therefore will not rip them apart.

The particular clusters which make it to the Earth travel a distance of astronomical unit (Au) near the Sun suffering all along a tidal acceleration
\begin{equation}
a_{\rm tidal}^{\rm (Sun)} \simeq \frac{G_N M_{\odot} R}{{\rm Au}^3} \,.
\end{equation}
However, the resulting fractional spreading in the single near-solar passage lasting a time $t= {\rm Au}/v_{\rm Virial}$ is $\delta R/R = a_{\rm tidal}^{\rm (Sun)} t^2/2R \simeq 6 \times 10^{-3}$, which is negligible.

The velocity $v_{\rm Virial}$ here and in the following is the typical galactic virial velocity of the clusters, which is  $\sim 300$ km/{sec}. The fractional gain in internal energy $m_{\rm DM} (\delta v)^2$ of any DM particle in a cluster in a collision with another cluster for a collision with impact parameter $b \sim R$ is:
\begin{eqnarray}
\label{eqn:deltav}
\frac{(\delta v)^2}{v_{\rm esc}^2}
 \ = \ \frac{v_{\rm esc}^2}{v_{\rm Virial}^2}
 \ \sim \ (10 ^{-18} - 10^{-22} ) \times f \,.
\end{eqnarray}
To show this we note that the velocity change is $\delta v = a_{\rm tidal} \delta t$, with the tidal acceleration
\begin{eqnarray}
a_{\rm tidal} = \frac{f G_N M_{\rm cls}}{R^2}
\end{eqnarray}
and $\delta t  \sim R/v_{\rm Virial}$.  Thus we find:
\begin{eqnarray}
(\delta v) ^2\ = \
\left[ \frac{f G_N M_{\rm cls}}{R^2}\frac{R}{v_{\rm Virial}} \right]^2
\ = \ \frac{v_{\rm esc}^4}{v_{\rm Virial}^2}
\end{eqnarray}
which is equivalent to Eq.~(\ref{eqn:deltav}). Since the imparted velocity changes $\delta v$ have random directions their effect adds up in quadrature and we need ${\cal N} \sim (10^{18} - 10^{22})\times f^{-1}$ collisions for ${\cal N} (\delta v)^2$ to become $v_{\rm esc}^2$ so that the cluster will be disrupted. For the optimal clusters of interest we have, by the very definition of these special clusters, that any cluster encounters another cluster about once every year. The $ (10^{18} - 10^{22}) \times f^{-1}$ collisions needed to destabilize the cluster then require more than the Hubble  time of $10^{10}$ years so long as we impose the following upper bound on the enhancement factor $f$ of the internal DM-DM interactions:
\begin{eqnarray}
f \ < \ 10^8 \,.
\end{eqnarray}

We next consider the effect of ordinary Milky Way stars. For simplicity we will assume  $2\times 10^{11}$ stars each of solar mass uniformly distributed over a disc of 15 kpc radius and  1 kpc thickness so that the total volume is $7.1 \times 10^{11} \, {\rm pc}^3$ and an average density of stars $n_{*} \sim 0.2 \, {\rm pc}^{-3} = 6.7 \times 10^{-57} \, {\rm cm}^{-3}$.

A cluster will be tidally disrupted in a single encounter at  impact parameter $l$ relative to the star if the resulting velocity gained $\delta v$ equals or exceeds the escape velocity from the cluster $v_{\rm esc} ^2 = f {G_N M_c}/{R_c}$.\footnote{Another necessary condition is that the tidal acceleration exceed the surface acceleration of the cluster so that the particles on the surface will be able to start flying out during the transit.  Specifically it is $l < R (M_*/M_{\rm cls})^{1/3}$ with $M_\ast$ the star mass. However, since the transit time is much shorter than the typical ``period'' of motion in the star those DM particles which started escaping will fall down back to form a slightly more excited ``puffed up'' cluster once the distance to the star will much exceeds the above value. To avoid this fall-back we need to impose the stronger requirement that during the relatively short transit time $\delta t = l/v_{\rm Virial}$ the imparted internal velocity to the DM particles will exceed the escape velocity  $v_{\rm esc}$ from the cluster.} Once the tidal acceleration exceeds the cluster surface gravity, i.e. $a_{\rm tidal} > g_ {\rm cls}$, we can use
\begin{eqnarray}
 \delta v \ = \ a_{\rm tidal} t_{\rm transit} \sim \frac{a_{\rm tidal} l}{v_{\rm Virial}}
\end{eqnarray}
for the velocity gain of individual DM particles. Using
\begin{eqnarray}
a_{\rm tidal} \ = \ \frac{G_N R M_*}{l^3}
\end{eqnarray}
for the tidal acceleration with the star mass $M_* \sim M_\odot$, we then have
\begin{eqnarray}
(\delta v)^2 \ = \ \left[ \frac{G_N R M_{\odot}}{l^2 v_{\rm Virial}} \right]^2 \,.
\end{eqnarray}
Comparing this with the escape velocity from the cluster $v_{\rm esc}^2 = f G_N M_{\rm cls}/R$ yields the condition for tidal disruption:
\begin{eqnarray}
\label{eqn:l4}
l^4 \leq \frac{1}{f}
\frac{G_N R^3 M_\odot}{v_{\rm Virial}^2} \frac{M_\odot}{M_{\rm cls}}.
\end{eqnarray}
The solar escape velocity $v_{\rm esc}^{(\odot)} =  \left[ G_N M_\odot / R_\odot \right]^{1/2}$ with the solar radius $R_\odot = 7 \times {10}^{10}$ cm is $v_{\rm esc}^{(\odot)} \simeq 600 \, {\rm km/sec} \simeq 2 v_{\rm Virial}$, and Eq.~(\ref{eqn:l4}) can be rewritten as:
\begin{eqnarray}
\label{eqn:yy}
l^4 \ \lesssim \ \frac{4 R^3 R_\odot M_\odot}{f M_{\rm cls}}
\ = \ \frac{R R_\odot M_\odot}{f(RE) \rho_{\rm DM}} \,,
\end{eqnarray}
where we used $M_{\rm cls} = \frac{4\pi}{3} R^3 E  \rho_{\rm DM}$, with $\rho_{\rm DM}$ the average un-enhanced DM density.

Over Hubble time $t_H$ the circular cross-sectional area  $\pi l^2$ with the
clump at its center sweeps a volume
\begin{eqnarray}
V_{\rm swept} \ = \ \pi l^2 v_{\rm Virial} t_{H} \,.
\end{eqnarray}
Naively this volume would  contains ${\cal N}_*= V_{\rm swept} n_\ast$ stars (with $n_\ast$ the star density) which could disrupt the cluster. However, the DM particles, particularly those with zip through the disk with velocities higher than $v_{\rm Virial} \simeq 300$ km/sec and are more readily detected, spend most of the time outside the disk in the much larger halo which is almost star free.  Thus the last estimate of the number ${\cal N}_*$ of critical encounters should be reduced by the ratio of the disk and halo volumes:
\begin{eqnarray}
r \ = \ V_{\rm disk}/V_{\rm halo}
\ \simeq \ \pi r_{\rm disk}^2 h_{\rm disk} / \frac{4\pi}{3} r_{\rm halo}^3 \,.
\end{eqnarray}
Using $r_{\rm disk} = 15$ kpc and $h_{\rm disk} = 1$ kpc and $r_{\rm halo} = 40$ kpc as effective values for disk radius, disk thickness and halo radius, we find $r = 3 \times 10^{-3}$. The actual number of cluster-star collisions then is
\begin{eqnarray}
r V_{\rm swept} n_* \ \simeq \ r \pi l^2 n_s \times 10^{25} \, {\rm cm} \,,
\end{eqnarray}
where we used $v_{\rm Virial} t_H \simeq 10^{25}$ cm. Using Eq.~(\ref{eqn:yy}), we then have:
\begin{eqnarray}
{\cal N}_* \lesssim
(10^{25} \, {\rm cm}) \times
r\pi \left[ \frac{R_c R_\odot M_\odot }{(10^{15}\, {\rm cm}) f \rho_{\rm DM}} \right]^{1/2} n_s
\end{eqnarray}
Using the values of $R_\odot$, $M_\odot$, the above $\rho_{\rm DM}$, and the stelar density  $n_* \sim 0.2 {\rm pc}^{-3} \simeq 6.7 \times 10^{-57} {\rm cm}^{-3}$ we find that for the ``large clusters'' of interest namely of sizes $10^{9} \, {\rm cm} < R < 10^{13} \, {\rm cm}$, ${\cal N}_*$ varies between $\lesssim 0.01 f^{-1/2}$ and $\lesssim f^{-1/2}$ at the lower and upper ends of the above range

The above discussion is clearly rather approximate. Thus the density of stars falls off with the distances $r$ and $h$ from the galactic center and the galactic disk and at the galactic center is significantly higher than the average $n_{\rm star} \sim 0.2 {\rm pc}^{-3}$ we used.  However, the motion of the DM particles inside the galaxy is rather chaotic as is likely to be the case when the galaxy being built up by successive mergers. Then collectively the DM particles will sample all regions of the galaxy (and halo) and justify our use of the average $n_{\rm star}$. There is a large enhancement of the  DM density towards the galactic center and the clusters therein are likely to be tidally disrupted. However the velocity of the DM particles  in the galactic center will tend to be much lower than $v_{\rm Virial}$. These DM clusters/particles will tend to stay in the inner regions. Since a) theses clusters/particles are not likely to visit us, and b) they are much less likely -- because of the lower recoil energies -- to be detected in the underground DM searches, this will not modify much the overall impact of the tidal disruption of clusters by stars.

We also did not consider the accumulative build up of internal kinetic energy of the DM particles inside a cluster by repeated collisions with stars smaller than the Sun which eventually can lead to the cluster's break-up. The stellar mass function  peaks at $\sim 0.6$ solar masses and so this may be only a mild effect. In general collisions with lower mass stars are less effective and  by pushing to a  higher common solar mass we enhanced the estimate of the overall tidal disruption effect due to stars, making it a conservative assumption.

Our estimates suggest that while just the pure self gravity of the smaller ($R < 10^9$ cm) clusters can overcome the disruption by stellar encounters, it may do it only marginally for the case of the larger $R > 10^{13}$ cm clusters of interest. This fact and the uncertainty in the above parameters suggests using the extra attractive DM-DM interactions, and $f \sim 10^4$ will suffice to evade any significant tidal disruption.

What is the effect of $G_N \to fG_N$ on the monochromaticity of the DM particles in a cluster in the $f=1$ (only gravity) case?  The resulting increase of the internal kinetic energy, proportional to $v_{esc}^2 \propto f$, generates a fractional deviation from the monochromatic kinetic energy of $1/2 m_{\rm DM} v^2$ (with $v \sim v_{\rm Virial}$) of all the DM particles inside the cluster:
\begin{eqnarray}
\frac{\Delta v}{v} \ \sim \ \frac{f^{1/2} v_{\rm esc}}{v}
\ \sim \ (10^{-11}-10^{-9}) \times f^{1/2}
\end{eqnarray}
where we used  $v_{esc}=  (4 \times 10^{-4} - 4 \times 10^{-2} )$ cm/sec relevant to the clusters of interest and $v \sim v_{\rm Virial} \sim 300$ km/sec for the motion of the cluster.  Thus so long as $f< 10^{8}$ the mono-chromatic nature of the DM particles in a cluster will be preserved.

\section{The issue of cluster formation}
\label{sec:formation}

\subsection{The case of small clusters/grains}
\label{sec:formation1}

The formation of grains/blobs of cold DM has been discussed by many authors. Specifically one invokes of some binding interactions in the dark sector, leading to novel forms of atomic or nuclear-like DM.

The new interactions required are relatively short range. Since the small clusters are not the main focus of this paper, we will not dwell on these here any further. We will see however in the next subsection that any formation of tiny grains/blobs of DM can help resolve the much tougher issue of forming the large clusters of interest.

\subsection{The case of large clusters/clouds}

The CMB data suggest primordial density fluctuations $\delta \rho/\rho \sim 10^{-5}$. The scales for which these fluctuations
grow under the influence of gravity alone forming autonomous structures, decoupled from the cosmological Hubble expansion, are
however limited by many considerations:
\begin{enumerate}
  \item A basic, causality, constraint is that only fluctuations which entered the horizon i.e. having sizes $R$ smaller than the horizon, can grow at time $t$.
  \item Ordinary matter structures do not grow prior to electron-proton recombination which occurs at CMB temperatures of 0.3 eV or red-shift of $z \sim 1000$. This need not apply to growth of structure in the cold DM particles of interest here, though analogous limits hold when we have light dark photons as  in mirror/twin analogs of the SM and in some other non-minimal dark sectors.
\item All fluctuations in both the ordinary and the dark sector can grow only logarithmically before the universe becomes matter dominated at $z \sim 10^4$, at which time fast linear growth can start.
  \item Another condition required for growth to start is that the potential wells due to the density fluctuation of the desired size $R$ and mass $M$ will be able to trap the DM particles having a kinetic energy of $m_{\rm DM}v_{\rm DM}^2/2$ (with $v_{\rm DM}$ the DM velocity at the time). This should happen already at the start of growth at time $t_{\rm start}$ or $z=z_{\rm start}$ before the density perturbations get enhanced. Only the mass excess inside the region of size $R$ with a positive density fluctuation:
\begin{eqnarray}
  M \left( \frac{\delta \rho}{\rho} \right) \sim 10^{-5}M
\end{eqnarray}
is pulling more cold DM into this region. Allowing for the enhanced $G_N \to f G_N$ gravity trapping requires that
\begin{eqnarray}
\label{eqn:trap}
\frac{f G_N M_{\rm cls}}{R} \frac{\delta \rho}{\rho} > v_{\rm DM}^2
\end{eqnarray}
\end{enumerate}

Even if the potential on the l.h.s. of Eq.~(\ref{eqn:trap}) is too weak to capture the typical cold DM with the average $v_{\rm DM}^2$ it still can
capture the fraction of DM particles whose velocities are low enough
so as to satisfy the condition in Eq.~(\ref{eqn:trap}).\footnote{This is reminiscent of the boiling off of the hotter atoms and keeping the colder atoms in shallow traps in preparing BEC systems.} This in turn may increase the effective trapping mass beyond $10^{-5} M$, capture a larger fraction of the DM etc. Thus we may have a weaker, easier to satisfy condition. For simplicity we will not follow this here and conservatively keep the stronger condition in Eq.~(\ref{eqn:trap}).

If all particles that at the start were within the desired radius $R$ stay trapped  and form the cluster, then the mass $M_{\rm cls}$ is the mass of
the cold DM particles contained therein:
\begin{eqnarray}
M_{\rm cls} = \frac{4\pi}{3} R^3 \rho(z_{\rm start})
\end{eqnarray}
at that time. For $\Omega_{\rm DM} \sim 0.2$, the cosmological DM density now is $0.2 \rho_{c} \simeq 0.6 \, {\rm keV}/{\rm cm}^3$. In the interim between $z_{\rm start}$  and now ($z=0$),  $\rho_{\rm DM}$ scales as $T^3 \sim 1/{(z+1)^3}$ and the cluster masses are :
\begin{eqnarray}
M_{\rm cls} =
\frac{4\pi}{3} R^3 (z_{\rm start}+1)^3 \Omega_{\rm DM} \rho_c     
\end{eqnarray}
Equating this to the present cluster mass
\begin{eqnarray}
M_{\rm cls} \simeq \frac{4\pi}{3} R^3 E \rho_{\rm DM} (0),
\end{eqnarray}
we find that, for $10^9 \, {\rm cm} < R < 10^{13}$ cm (with $10^6>E >10^2$),
\begin{eqnarray}
4 \times 10^2 < z_{\rm start} < 8\times 10^3 \,,
\end{eqnarray}
amusingly this is consistent (though only marginally at the upper end)
with condition (3) of matter domination at the start of fast growth.

Returning to Eq.~(\ref{eqn:trap}) we find that the velocity $v_{\rm DM}$ of DM
particles which stayed in equilibrium with some ``dark radiation'' of
temperature $T'\sim T$ until $z=z_{\rm start}$ is too large to allow the
condition (\ref{eqn:trap}) to be satisfied for reasonable DM mass $m_{\rm DM}$. We will therefore assume that the DM does not couple to any radiation
or that it decouples very early when $T' \sim T$ are similar  to the mass
$m_{\rm DM}$. In this case once the DM becomes non-relativistic, say with velocity of $v \sim c/3$, at $z_ {\rm initial} \sim m_{\rm DM} /T(0)$ its momentum $p$ and velocity $v \sim p/{m_{\rm DM}}$ get red-shifted in proportion to $(z+1)$ so that
at $z_{\rm start}$ we have
\begin{eqnarray}
v_{\rm DM}^2 \sim \left[ c \times \frac{z_{\rm start}}{z_{\rm intial}} \right]^2
\simeq 60 \left[ \frac{m_{\rm DM}}{\rm GeV} \right]^2 {\rm cm}^2/{\rm sec}^2
\end{eqnarray}
and the condition (\ref{eqn:trap}) is fulfilled if the DM mass is large enough:
\begin{eqnarray}
m_{\rm DM} > 10^{7} \, {\rm GeV} \times f^{-1/2}\quad \text{and} \quad
m_{\rm DM} > 10^5 \, {\rm GeV} \times f^{-1/2}
\end{eqnarray}
for $R = 10^9$ cm and $R =10^{13}$ cm, respectively.

Using the maximal enhancement factor $f=10^8$ allowed by demanding stability against mutual cluster-cluster collisions,  the required DM masses $m_{\rm DM}$  need not exceed $10^3$ GeV. This is consistent with an upper bound on $m_{\rm DM}$~\cite{Griest:1989wd}. This bound applies for elementary (non-composite) DM, which was early on in thermal equilibrium. It is required in order to allow sufficiently fast DM annihilation so as to leave the correct  DM relic density.

This possible high mass DM can be avoided by
reducing the random velocities hindering the start of optimal cluster formation.
Thus let us assume that in some early epoch the temperature $T$ in the dark sector has dropped bellow the typical (nuclear-like or other) binding energy of a DM in a nucleus or more generally some dark blob. Formation of such blobs will then start. Initially it is via $\chi + \chi \to \chi^2$ composites and later also $\chi^n + \chi^m \to \chi^{n+m}$ will proceed. In general dark photons, not indicated in the above, carry the excess energy. Once most DM is inside blobs, the growth of the blobs naturally stops at some critical number ${\cal N}$ of elementary $\chi$ in the blob. These new blobs of mass $M \sim {\cal N} m_{\rm DM}$  will then be the new effective DM. The randomly directed momenta $p_i$ of the ${\cal N}$ DM particles constituting the blob add in quadrature $P^2 = p_1^2 + p_2^2 + \cdots + p_N^2$  so that the squared velocity of the blob
\begin{eqnarray}
v_{blob}^ 2 \ = \
\left(\frac{P}{M}\right)^2  \ = \ \frac{v^2}{\cal N}
\end{eqnarray}
with $v$ the average velocity of the initial DM particles. The much smaller velocities of the DM blobs will then accelerate the formation of the clusters/clouds of interest. The latter will then be made of the above DM blobs. In principle ${\cal N}$ and the mass of the blob can be ``dialed'' to be very large so long as the latter is smaller than the mass of the critical grain introduced in Eq.~(\ref{eqn:criticalmass}), as otherwise we will encounter the grains only once in a period longer than one year. The original much lighter DM particles freeze out much before the blob/new-heavy-DM formation described above, and no conflict with the Greist-Kamionokowski bound on DM mass~\cite{Griest:1989wd} will arise.

We need to address yet another severe obstacle to the formation of the relatively small clusters.  The assumed $z_{\rm start} \sim 10^4$ corresponds to times $t_{\rm start} \sim 10^{-8} t_{H}$ and horizon sizes of $ct_{\rm start} \sim 10^{20}$ cm. Thus perturbations on scales far larger than the scales $R = (10^9 - 10^{13})$ cm of interest, have by then entered the horizon and can start growing.  The gravitational potential wells generated by these larger structures will  accelerate the DM particles  to velocities far exceeding the values reached via unhindered red-shifting.  This excludes the possibility of generating our ``optimal'' clusters by gravity only. Indeed detailed many-body simulations of early structure allow -- even in the most optimistic case -- the formation of micro-halos down to Earth masses~\cite{Goerdt:2006hp}, which still are many orders of magnitude above those of the desired values.

DM could collapse and form the clusters of interest if their formation period were shifted to much earlier times before the build-up of larger structures and attendant larger velocities of DM particles. Introducing the extra attractive force between the DM particles can help achieving the optimal cluster formation by:
\begin{enumerate}
  \item Decoupling of the force generating the clusters from the force of gravity which controls the Hubble expansion will allow fast growth of the smaller structures of interest {\it before} $z \sim 10^4$ when the universe ceases to be radiation dominated and the large ordinary gravity generated clusters can start growing.

  \item Choosing the range of the new attractive force to be  the size  of cluster $10^9 < R < 10^{13}$ cm of interest selectively enhances the formation of the corresponding clusters. They will have a head start (due to point 1 above) relative to the formation of the larger pure gravity induced structures and their growth will be  faster due to the stronger effective gravity which drives this growth.
\end{enumerate}

The new force could be generated by the exchange of a new scalar $s$ of mass:
\begin{eqnarray}
m_s \sim R^{-1} \sim (10^{-18}- 10^{-14}) \, {\rm eV} \,.
\end{eqnarray}
Even a tiny Yukawa coupling of this scalar to the DM particles:
\begin{eqnarray}
g^2 \ = \
f G_N m_{\rm DM}^2= f \times 10^{-38} \left( \frac{m_{\rm DM}}{\rm GeV} \right)^2
\end{eqnarray}
would enhance by a factor of $f$ the effective Newton constant $G_N$ which governs the mutual attraction of DM particles. Clearly the $s$ exchange would be  coherent and additive over the tiny size of the putative blobs whose formation could decrease the DM velocity by a large factor of ${\cal N}_{\rm DM}$. Consequently also the blob-blob interaction due to its exchange scales with the number ${\cal N}_{\rm DM}$ of elementary DM particles in it and with the mass $M = {\cal N}_{\rm DM} m_{\rm DM}$, just like gravity. While  detailed elaboration of the last scenario is far beyond the scope of this paper, the above discussion suggest that with some moderate amount of new physics formation of the ``large clusters'' of interest may be possible.

Finally we would like to mention another interesting approach to the generation of relatively small structures. It utilizes a modified cosmology where an earlier matter dominated phase occurs before the ordinary radiation dominated stage.  This scenario has been recently considered by several authors~\cite{Erickcek:2011us, Blinov:2019jqc} and requires the existence of very massive particles, which are however rather long-lived and decay prior to nucleosynthesis allowing the decay products to get into thermal equilibrium. The  heavy DM dominated era can last long enough so that the ordinary stable cold DM particles can start collapsing  early on, making structures which are far smaller than those arising in ordinary cosmology.

\section{Conclusion}
\label{sec:conclusion}

In this paper we analyse the possible impact of DM clustering at various scales on direct DM searches. Our main message is that looking for time correlations of events in different large underground detectors searching for DM may be of enormous value. Such time correlations between events in completely independent experiments located in different continents cannot be accidental. They do arise if DM clusters on specific scales ranging in sizes between Earth size to $10^4$ times Earth size, with appropriate density contrast, if the clusters contain an appreciable fraction of all DM.

Finding such time correlations requires a minimal effort of different experimental groups. All that needs to be done is simply to compare the times of the ``good'' and ``moderately good'' candidates for WIMP-nuclear scattering seen in the different detectors. There is no excuse for not doing this before embarking on yet further huge spending on one much larger detector. The effort involved in the extra analysis pales in comparison with the huge resources, efforts and ingenuity invested in order to obtain these rare events.

What we are offering here is in effect a ``very bright lamp'' under which DM -- an important key to understanding the universe -- may be found. If it is indeed found there the theoretical particle physics community will certainly find how the key got there, namely produce a convincing complete model for their stability and for their formation. Still to indicate that such scenarios are viable we discuss in some detail in Section~\ref{sec:stability} the issue of the stability of the large clusters under tidal disruption by the galaxy, the Sun, other Milky Way stars and in  mutual cluster-cluster collisions.

We also addressed in depth
the issue of the cluster formation. We recalled the various reasons why such relatively small clusters will not form under the influence of gravity alone and pointed out that some new physics specific to the dark sector may allow achieving this. In particular we considered  DM-DM attractive force of range comparable to the radius of the DM clusters of interest.  The strength of the new attraction is limited by the need  to maintain cluster stability under the many cluster-cluster collisions to be less than $\sim 10^8$ times that of gravity. This may still allow the small structures in the dark sector to start growing earlier and faster than the competing larger structures due to gravity alone. Another interesting point is that formation of small nuclei/blobs  via some short range interaction tends to drastically cool the DM (which then consists of these blobs)  and facilitates  structure growth.

In addition to the main new  message concerning the large clusters/clouds, we reviewed the case of  small clusters/blobs/grains which, when reaching a critical mass,  leave unique signatures in each of the DM detectors.

Searches of non-WIMP cold DM, such as those looking for resonant axion-photon conversion in magnetic field done at widely separated facilities, can also benefit from the approximate coincidences due to optimal clustering, just as the Xenon experiments which we focused on.
Interestingly also searches of monochromatic microwave photons from axion conversion in the magnetic fields of white dwarfs or neutron stars will be modified by formations of axion clusters. Because of the astrophysical context the clusters can be on scales or $R$ and $E$ (or $R$ and $M_{\rm cls}$) values very different from what was considered here. All of the above modify the expected signals in the same common, characteristic way: the emission will tend to be much rarer as only a fraction $1/E$ of the neutron stars or white dwarfs will overlap a cluster at any given time. Yet the emission in these cases will be $E$ times stronger, leading to more dramatic signals and higher sensitivity.

\section*{Acknowledgements}
The authors would like to thank Jim Buckley, Ram Cowsik, Jordan Goodman, Carter Hall, Zohar Nussinov, Robert Shrock, Yushin Tsai, Philip Mannheim and Kuver Sinha for helpful discussions. We would also thank the anonymous referee
whose inquires and suggestions led to a clearer and better paper. The work of Y.Z. is supported by the US Department of Energy under grant No. DE-SC0017987. Y.Z. is grateful to the Center for Future High Energy Physics, Institute of High Energy Physics, CAS and the Center for High Energy Physics, Peking University for the hospitality and local support where part of the work was done.


\begin{thebibliography}{99}

\bibitem{Bird:2016dcv}
  S.~Bird, I.~Cholis, J.~B.~Munoz, Y.~Ali-Haimoud, M.~Kamionkowski, E.~D.~Kovetz, A.~Raccanelli and A.~G.~Riess,
  Phys.\ Rev.\ Lett.\  {\bf 116}, no. 20, 201301 (2016)
  [arXiv:1603.00464 [astro-ph.CO]].

\bibitem{Hui:2016ltb}
  L.~Hui, J.~P.~Ostriker, S.~Tremaine and E.~Witten,
  Phys.\ Rev.\ D {\bf 95}, no. 4, 043541 (2017)
  [arXiv:1610.08297 [astro-ph.CO]].

\bibitem{Witten:1984rs}
  E.~Witten,
  Phys.\ Rev.\ D {\bf 30}, 272 (1984).

\bibitem{Bai:2018dxf}
  Y.~Bai, A.~J.~Long and S.~Lu,
  Phys.\ Rev.\ D {\bf 99}, 055047 (2019)
  [arXiv:1810.04360 [hep-ph]].

\bibitem{Ge:2019voa}
  S.~Ge, K.~Lawson and A.~Zhitnitsky,
  arXiv:1903.05090 [hep-ph].

\bibitem{Grabowska:2018lnd}
  D.~M.~Grabowska, T.~Melia and S.~Rajendran,
  arXiv:1807.03788.

\bibitem{Feng:2010gw}
  J.~L.~Feng,
  Ann.\ Rev.\ Astron.\ Astrophys.\  {\bf 48}, 495 (2010)
  [arXiv:1003.0904 [astro-ph.CO]].

\bibitem{Nussinov:1985xr}
  S.~Nussinov,
  Phys.\ Lett.\  {\bf 165B}, 55 (1985).

\bibitem{Kaplan:2009ag}
  D.~E.~Kaplan, M.~A.~Luty and K.~M.~Zurek,
  Phys.\ Rev.\ D {\bf 79}, 115016 (2009)
  [arXiv:0901.4117 [hep-ph]].

\bibitem{Spergel:1999mh}
  D.~N.~Spergel and P.~J.~Steinhardt,
  Phys.\ Rev.\ Lett.\  {\bf 84}, 3760 (2000)
  [astro-ph/9909386].

\bibitem{Hochberg:2014dra}
  Y.~Hochberg, E.~Kuflik, T.~Volansky and J.~G.~Wacker,
  Phys.\ Rev.\ Lett.\  {\bf 113}, 171301 (2014)
  [arXiv:1402.5143 [hep-ph]].

\bibitem{Berezhiani:2003xm}
  Z.~Berezhiani,
  Int.\ J.\ Mod.\ Phys.\ A {\bf 19}, 3775 (2004)
  [hep-ph/0312335].

\bibitem{Foot:2014mia}
  R.~Foot,
  Int.\ J.\ Mod.\ Phys.\ A {\bf 29}, 1430013 (2014)
  [arXiv:1401.3965 [astro-ph.CO]].

\bibitem{Berezhiani:1995am}
  Z.~G.~Berezhiani, A.~D.~Dolgov and R.~N.~Mohapatra,
  Phys.\ Lett.\ B {\bf 375}, 26 (1996)
  [hep-ph/9511221].

\bibitem{Chacko:2005pe}
  Z.~Chacko, H.~S.~Goh and R.~Harnik,
  Phys.\ Rev.\ Lett.\  {\bf 96}, 231802 (2006)
  [hep-ph/0506256].

\bibitem{Chacko:2005un}
  Z.~Chacko, H.~S.~Goh and R.~Harnik,
  JHEP {\bf 0601}, 108 (2006)
  [hep-ph/0512088].

\bibitem{Abercrombie:2015wmb}
  D.~Abercrombie {\it et al.},
  arXiv:1507.00966 [hep-ex].

\bibitem{Ackermann:2015zua}
  M.~Ackermann {\it et al.} [Fermi-LAT Collaboration],
  Phys.\ Rev.\ Lett.\  {\bf 115}, no. 23, 231301 (2015)
  [arXiv:1503.02641 [astro-ph.HE]].

\bibitem{Aartsen:2012kia}
  M.~G.~Aartsen {\it et al.} [IceCube Collaboration],
  Phys.\ Rev.\ Lett.\  {\bf 110}, no. 13, 131302 (2013)
  [arXiv:1212.4097 [astro-ph.HE]].

\bibitem{Tan:2016zwf}
  A.~Tan {\it et al.} [PandaX-II Collaboration],
  Phys.\ Rev.\ Lett.\  {\bf 117}, no. 12, 121303 (2016)
  [arXiv:1607.07400 [hep-ex]].

\bibitem{Cui:2017nnn}
  X.~Cui {\it et al.} [PandaX-II Collaboration],
  Phys.\ Rev.\ Lett.\  {\bf 119}, no. 18, 181302 (2017)
  [arXiv:1708.06917].

\bibitem{Aprile:2017iyp}
  E.~Aprile {\it et al.} [XENON Collaboration],
  Phys.\ Rev.\ Lett.\  {\bf 119}, no. 18, 181301 (2017)
  [arXiv:1705.06655].

\bibitem{Aprile:2018dbl}
  E.~Aprile {\it et al.} [XENON Collaboration],
  Phys.\ Rev.\ Lett.\  {\bf 121}, no. 11, 111302 (2018)
  [arXiv:1805.12562 [astro-ph.CO]].

\bibitem{Akerib:2016vxi}
  D.~S.~Akerib {\it et al.} [LUX Collaboration],
  Phys.\ Rev.\ Lett.\  {\bf 118}, no. 2, 021303 (2017)
  [arXiv:1608.07648].

\bibitem{Essig:2016crl}
  R.~Essig, J.~Mardon, O.~Slone and T.~Volansky,
  Phys.\ Rev.\ D {\bf 95}, no. 5, 056011 (2017)
  [arXiv:1608.02940 [hep-ph]].

\bibitem{Budnik:2017sbu}
  R.~Budnik, O.~Chesnovsky, O.~Slone and T.~Volansky,
  Phys.\ Lett.\ B {\bf 782}, 242 (2018)
  [arXiv:1705.03016 [hep-ph]].

\bibitem{Knapen:2017ekk}
  S.~Knapen, T.~Lin, M.~Pyle and K.~M.~Zurek,
  Phys.\ Lett.\ B {\bf 785}, 386 (2018)
  [arXiv:1712.06598 [hep-ph]].

\bibitem{Strassler:2006im}
  M.~J.~Strassler and K.~M.~Zurek,
  Phys.\ Lett.\ B {\bf 651}, 374 (2007)
  [hep-ph/0604261].

\bibitem{Randall:2014lxa}
  L.~Randall and M.~Reece,
  Phys.\ Rev.\ Lett.\  {\bf 112}, 161301 (2014)
  [arXiv:1403.0576 [astro-ph.GA]].

\bibitem{Grossman:2017qzw}
  Y.~Grossman, R.~Harnik, O.~Telem and Y.~Zhang,
  arXiv:1712.00455 [hep-ph].

\bibitem{Goerdt:2006hp}
  T.~Goerdt, O.~Y.~Gnedin, B.~Moore, J.~Diemand and J.~Stadel,
  Mon.\ Not.\ Roy.\ Astron.\ Soc.\  {\bf 375}, 191 (2007)
  [astro-ph/0608495].

\bibitem{Baum:2018tfw}
  S.~Baum, A.~K.~Drukier, K.~Freese, M.~G\'{o}rski and P.~Stengel,
  arXiv:1806.05991.












\bibitem{Krnjaic:2014xza}
  G.~Krnjaic and K.~Sigurdson,
  Phys.\ Lett.\ B {\bf 751}, 464 (2015)
  [arXiv:1406.1171 [hep-ph]].

\bibitem{Hardy:2014mqa}
  E.~Hardy, R.~Lasenby, J.~March-Russell and S.~M.~West,
  JHEP {\bf 1506}, 011 (2015)
  [arXiv:1411.3739 [hep-ph]].

\bibitem{Coskuner:2018are}
  A.~Coskuner, D.~M.~Grabowska, S.~Knapen and K.~M.~Zurek,
  arXiv:1812.07573 [hep-ph].

\bibitem{Gelmini:2002ez}
  G.~Gelmini, A.~Kusenko and S.~Nussinov,
  Phys.\ Rev.\ Lett.\  {\bf 89}, 101302 (2002)
  [hep-ph/0203179].

\bibitem{DeRujula:1984axn}
  A.~De Rujula and S.~L.~Glashow,
  Nature {\bf 312}, 734 (1984).

\bibitem{Hall}
  Carter Hall, private communication.

\bibitem{Diemand:2009bm}
  J.~Diemand and B.~Moore,
  Adv.\ Sci.\ Lett.\  {\bf 4}, 297 (2011)
  [arXiv:0906.4340 [astro-ph.CO]].

\bibitem{Arvanitaki:2019rax}
  A.~Arvanitaki, S.~Dimopoulos, M.~Galanis, L.~Lehner, J.~O.~Thompson and K.~Van Tilburg,
  arXiv:1909.11665 [astro-ph.CO].

\bibitem{Erickcek:2011us}
  A.~L.~Erickcek and K.~Sigurdson,
  Phys.\ Rev.\ D {\bf 84}, 083503 (2011)
  [arXiv:1106.0536 [astro-ph.CO]].

\bibitem{Blinov:2019jqc}
  N.~Blinov, M.~J.~Dolan and P.~Draper,
  arXiv:1911.07853 [astro-ph.CO].

\bibitem{Griest:1989wd}
  K.~Griest and M.~Kamionkowski,
  Phys.\ Rev.\ Lett.\  {\bf 64}, 615 (1990).

\bibitem{Nussinov:2014qva}
  S.~Nussinov,
  arXiv:1408.1157.





\end{thebibliography}
\end{document}